\documentclass{article}

\usepackage[preprint]{neurips_2025}

\usepackage[utf8]{inputenc} 
\usepackage[T1]{fontenc}    
\usepackage{hyperref}       
\usepackage{url}            
\usepackage{booktabs}       
\usepackage{amsfonts}       
\usepackage{nicefrac}       
\usepackage{microtype}      
\usepackage{xcolor}         
\usepackage{amsmath,amssymb,amsfonts}
\usepackage{graphicx}
\usepackage{textcomp}
\usepackage{xcolor}
\def\BibTeX{{\rm B\kern-.05em{\sc i\kern-.025em b}\kern-.08em
    T\kern-.1667em\lower.7ex\hbox{E}\kern-.125emX}}

\usepackage{booktabs} 
\usepackage{import}
\usepackage{graphicx}
\usepackage{epstopdf}
\usepackage{amsmath}
\usepackage{subfigure}
\usepackage{diagbox}
\usepackage{float}
\graphicspath{{Fig/}}
\usepackage{booktabs}
\usepackage{algorithm}
\usepackage{algorithmicx}
\usepackage{algpseudocode}
\usepackage{paralist}
\usepackage{multirow}
\usepackage{graphicx}
\usepackage{subfigure}
\usepackage{color}
\usepackage{hhline}
\usepackage{xcolor}
\usepackage{array}
\usepackage{diagbox}
\usepackage{hyperref}
\usepackage{verbatim}
\usepackage{colortbl}
\usepackage{upgreek}

\usepackage{tikz}
\usetikzlibrary{arrows.meta,positioning,shapes.geometric,calc,fit,backgrounds}
\usepackage{enumitem}
\usepackage{gensymb} 
\usepackage{amsmath,bm}
\usepackage{amssymb}
\usepackage{pifont}
\usepackage{filecontents}
\algdef{SE}[DOWHILE]{Do}{doWhile}{\algorithmicdo}[1]{\algorithmicwhile\ #1}%
\usepackage{amsmath, amssymb, amsthm}
\usepackage{algorithm}

\definecolor{deep_pink}{HTML}{D97692}

\title{\Large \bf MPAC: A Multi-Principal Agent Coordination Protocol\\
for Interoperable Multi-Agent Collaboration}
\author{%
  Kaiyang Qian  \\
  aistatus.cc \\
  University of Colorado Denver (CU Denver)\\
  Denver, CO 80204 \\
  \texttt{kaiyang.2.qian@ucdenver.edu} \\
  \And
  Xinmin Fang \\
  aistatus.cc \\
  University of Colorado Denver (CU Denver)\\
  Denver, CO 80204 \\
  \texttt{xinmin.fang@ucdenver.edu} \\
  \AND
  Zhengxiong Li \\
  aistatus.cc \\
  University of Colorado Denver (CU Denver)\\
  Denver, CO 80204 \\
  \texttt{zhengxiong.li@ucdenver.edu} \\
}

\begin{document}

\maketitle
\begin{abstract}
The AI agent ecosystem has converged on two protocols: the Model Context Protocol (MCP) for tool invocation and Agent-to-Agent (A2A) for single-principal task delegation. Both assume a \emph{single controlling principal}---one person or organization that owns and trusts every agent in the system. When independent principals' agents must coordinate over shared state---two engineers' coding agents editing the same repository, family members' agents planning a shared trip, agents from different organizations negotiating a joint decision---neither protocol applies, and coordination collapses to ad-hoc chat, manual merging, or silent overwrites. We present \textbf{MPAC} (Multi-Principal Agent Coordination Protocol), an application-layer protocol that fills this gap with explicit coordination semantics across five logical layers---Session, Intent, Operation, Conflict, and Governance. MPAC makes intent declaration a precondition for action, represents conflicts as first-class structured objects rather than silent side-effects, and supports human-in-the-loop arbitration through a pluggable governance layer. The specification defines 21 message types, three state machines with normative transition tables, Lamport-clock watermarking for causal ordering, two execution models (pre-commit and post-commit), three security profiles, and an optimistic-concurrency-control mechanism for shared state. We release two interoperable reference implementations (Python, 122 tests; TypeScript, 101 tests; 66 adversarial enforcement tests total), a machine-readable JSON Schema suite covering all 21 message types, and seven live multi-agent demos spanning code editing, consumer trip planning, pre-commit authorization with fault recovery, and multi-level conflict escalation. A controlled three-agent cross-module code review benchmark reports a \textbf{95\% reduction in coordination overhead} (68.65\,s $\rightarrow$ 3.02\,s) and a \textbf{4.8$\times$ wall-clock speedup} (131.76\,s $\rightarrow$ 27.38\,s) under MPAC compared to a serialized human-mediated baseline, with \emph{per-agent decision time preserved} (63.11\,s $\rightarrow$ 57.13\,s)---demonstrating that the speedup comes from eliminating coordination waits, not from compressing model calls. The full specification, reference implementations, test suites, and demo transcripts are released as open source.
\end{abstract}

\vspace{-0.5em}

\section{Introduction}
\label{sec:intro}

The artificial intelligence (AI) agent ecosystem is growing fast around two complementary protocols. The \textbf{Model Context Protocol (MCP)} \cite{anthropic2024mcp} standardizes how a single agent discovers and invokes external tools: file systems, databases, SaaS APIs, and so on. Its success is evident in the rapid proliferation of MCP servers for commercial services. The \textbf{Agent-to-Agent (A2A) protocol} \cite{google2024a2a}, introduced shortly after, standardizes how a single orchestrator agent delegates sub-tasks to worker agents under its authority, with a shared notion of ``task'' and ``artifact.'' Both protocols share a foundational assumption: there is \emph{one principal}---one person, team, or organization---that owns, trusts, and is accountable for every agent participating in the interaction.

This single-principal assumption is becoming a bottleneck. Consider a concrete and entirely realistic scenario. Alice and Bob are two engineers on the same software team. Alice runs a security-focused coding agent; Bob runs a code-quality coding agent. Both agents have full edit access to the shared repository. Alice's agent decides to patch a token expiry validation bug in \texttt{auth.py} and \texttt{auth\_middleware.py}. Bob's agent, operating in parallel, decides to refactor authentication logic to remove duplication---across \texttt{auth.py}, \texttt{auth\_middleware.py}, and \texttt{models.py}. Neither agent reports to the other. Neither principal has authority over the other's agent. There is no orchestrator. MCP is silent on the problem (it only mediates tool calls, not inter-agent coordination). A2A is silent as well (no single principal can legitimately tell both agents what to do). The agents proceed independently and either produce a merge conflict, silently overwrite each other's work, or---worst of all---produce a plausible-looking but semantically inconsistent merged result.

The same pattern recurs outside software: family members' agents negotiating a shared vacation itinerary, agents from different legal teams drafting a joint contract, smart-home agents belonging to different residents resolving HVAC preferences, and agents from competing organizations negotiating a supply-chain allocation. Each of these involves \emph{multiple independent principals} whose agents must coordinate over shared state, but none has a single orchestrator with authority over all participants. The coordination layer between independent principals and their agents is simply missing from today's protocol stack.

This paper presents \textbf{MPAC (Multi-Principal Agent Coordination Protocol)}, an application-layer protocol designed to fill this gap. MPAC is not a replacement for MCP or A2A---it is a complementary layer. MCP tells an agent how to call a tool; A2A tells an orchestrator how to delegate work; MPAC tells agents from different principals how to \emph{coordinate} when no orchestrator exists.

\paragraph{Contributions.}
\begin{enumerate}[leftmargin=1.3em, itemsep=1pt, topsep=2pt]
    \item \textbf{Problem framing.} We articulate the multi-principal coordination gap as a distinct protocol-layer problem, separate from tool invocation (MCP) and orchestrated delegation (A2A), and enumerate the concrete guarantees a solution must provide.
    \item \textbf{Protocol design.} We present MPAC, a five-layer protocol (Session, Intent, Operation, Conflict, Governance) with 21 message types, 3 state machines with normative transition tables, Lamport-clock causal watermarking, two execution models (pre-commit / post-commit), three security profiles (open / authenticated / verified), and explicit optimistic concurrency control on shared state. The design crystallizes a set of principles---\emph{intent before action}, \emph{attributable actions}, \emph{structured conflict}, and \emph{human-governed resolution}---into machine-enforceable wire semantics.
    \item \textbf{Reference implementations and interoperability.} We release two fully interoperable reference implementations, one in Python ($\sim$3{,}500 LOC, 122 tests) and one in TypeScript ($\sim$2{,}900 LOC, 101 tests), along with 66 adversarial tests specifically targeting enforcement bypass. A cross-language interop harness exchanges 14 messages bidirectionally with zero wire-format deviation.
    \item \textbf{Open JSON Schema suite.} All 21 message types and 4 shared object types are defined as JSON Schema (Draft 2020-12) with an envelope schema that dispatches payload validation by \texttt{message\_type} via \texttt{if/then} constraints, making third-party wire compatibility achievable by schema validation alone.
    \item \textbf{Empirical validation.} We run seven live multi-agent scenarios using Claude as the agent backend, covering: (i) concurrent code editing with optimistic concurrency control, (ii) consumer trip planning with multi-principal task-set negotiation, (iii) a controlled overhead-comparison benchmark, (iv) pre-commit authorization with agent fault recovery, and (v) multi-level conflict escalation with arbiter resolution. The overhead benchmark records a 95\% reduction in coordination overhead and a 4.8$\times$ wall-clock speedup while holding per-agent decision time roughly constant.
\end{enumerate}

The rest of the paper is organized as follows. Section~\ref{sec:problem} formalizes the multi-principal coordination problem and contrasts it with adjacent protocols. Section~\ref{sec:design} presents MPAC's design goals, non-goals, and shared principles. Section~\ref{sec:model} describes the protocol model and the five coordination layers. Section~\ref{sec:messages} enumerates the 21 message types and three state machines. Section~\ref{sec:security} covers security profiles, authorization, and governance. Section~\ref{sec:refimpl} describes the reference implementations and their adversarial test regime. Section~\ref{sec:eval} reports empirical results from the live Claude agent scenarios. Section~\ref{sec:related} surveys related work. Section~\ref{sec:limitations} honestly discusses limitations and threats to the empirical claims. Section~\ref{sec:conclusion} concludes.

\section{The Multi-Principal Coordination Problem}
\label{sec:problem}

We now describe the problem MPAC is trying to solve and make precise why existing tools do not solve it.

\subsection{What ``Multi-Principal'' Means}

A \textbf{principal} is an entity on whose behalf an agent acts and to whom the agent is accountable: a human user, a team, an organization, or an automated system with delegated authority. A session is \emph{single-principal} if every participating agent is accountable to the same principal (possibly indirectly, through an orchestrator). It is \emph{multi-principal} if two or more distinct, independent principals have at least one agent each, and no one principal has authority over the others' agents.

Multi-principal coordination is qualitatively different from single-principal coordination in three ways.

\paragraph{No unified decision-maker.} In a single-principal system, ambiguity or conflict can always be escalated to the principal and resolved by fiat. In a multi-principal system, no such fallback exists. If Alice's agent and Bob's agent disagree, there is no single human or meta-agent with the authority to simply decide. Resolution must either come from mutual agreement between principals, from a pre-agreed arbitration policy, or from escalation to an agreed arbiter---all of which need to be first-class protocol features, not implementation details.

\paragraph{Limited trust.} In single-principal systems, agents share a trust domain: they can assume other agents are correct, well-intentioned, and following the same policies. In multi-principal systems, agents from different principals may have different goals, different safety policies, different risk tolerances, and (in adversarial cases) different incentives to misreport state. The protocol must be defensible against adversarial or buggy participants, not just correct ones.

\paragraph{Auditability across boundaries.} In single-principal systems, audit logs live inside one organization. In multi-principal systems, every consequential action must be attributable across organizational boundaries---so that after the fact, each principal can independently verify what happened, who decided what, and based on what causal context.

\subsection{Concrete Scenarios}

To ground the discussion, we describe four scenarios that recur across domains and that all require multi-principal coordination. The same protocol primitives should handle all of them.

\begin{enumerate}[leftmargin=1.3em, itemsep=2pt, topsep=2pt]
    \item \textbf{Shared codebase, independent engineers.} Two or more engineers' coding agents independently select tasks from a backlog and edit a shared repository. Scope overlaps (same file, same function) must be detected before work begins, and optimistic concurrency control must be enforced at commit time.
    \item \textbf{Family trip planning.} Three family members' agents jointly plan a 5-day trip, each advocating for its principal's preferences (camping vs.~boutique hotel, theme park vs.~cultural workshop). They must negotiate overlapping claims on itinerary days and budget categories, and commit the resulting plan atomically.
    \item \textbf{Cross-organizational document drafting.} Legal agents from two counter-parties draft sections of a shared contract. Neither team can unilaterally overwrite the other's text, and disputed clauses must be escalated to human lawyers.
    \item \textbf{Multi-tenant resource allocation.} Agents from competing teams in a shared compute cluster negotiate job priorities. Conflicts must be surfaced to human administrators with full causal context.
\end{enumerate}

All four share the same structural requirements: (a) agents must declare \emph{what they plan to do} before doing it; (b) the system must automatically detect overlapping or contradictory claims; (c) conflicts must be surfaced as structured, attributable objects rather than silent overwrites; (d) resolution paths must include human override; and (e) all of the above must be auditable by each principal independently.

\subsection{Why Existing Protocols Do Not Suffice}

\paragraph{MCP is about tools, not coordination.} MCP standardizes how an agent discovers and calls tools exposed by an external server. It is deliberately silent on what happens when two agents call the same tool at the same time, on how agents learn of each other's existence, and on how mutually inconsistent actions are resolved. MCP is the right protocol for single-agent-to-tool interaction; it is the wrong layer for agent-to-agent coordination.

\paragraph{A2A is single-principal by design.} A2A assumes an orchestrator that delegates tasks to worker agents. The orchestrator owns the authority chain: it decides who does what, receives all artifacts, and is the point of resolution for any ambiguity. There is no notion in A2A of two independent orchestrators meeting as peers and negotiating a shared outcome. Extending A2A with ``just let the orchestrators talk to each other'' recreates the problem MPAC solves one level up---now there is no meta-orchestrator to decide between them.

\paragraph{Message queues and locks are too low-level.} One could imagine building multi-principal coordination on top of a message queue or a distributed lock. But these primitives address the \emph{mechanism} of serialization, not the \emph{semantics} of intent, conflict, or governance. They provide no standard for how an agent should announce a planned change, how overlapping scopes are reported, how disputes escalate, or how human override is integrated. Every application would reinvent these semantics incompatibly. MPAC's goal is to provide the semantic layer once, so that heterogeneous agent systems can interoperate.

\paragraph{CRDTs and OT address state, not coordination.} Conflict-free replicated data types (CRDTs) \cite{shapiro2011crdt} and operational transformation (OT) \cite{ellis1989concurrency} provide principled ways to merge concurrent edits to shared data. They are orthogonal to MPAC: they address \emph{how} to reconcile divergent state once a conflict has occurred, whereas MPAC addresses \emph{whether} the agents should have made those edits in the first place, \emph{who} gets to decide when preferences collide, and \emph{how} the decision is audited. MPAC can be layered on top of a CRDT-backed shared store without modification; the two solve complementary problems.

\paragraph{Agent frameworks are libraries, not protocols.} Frameworks such as LangGraph \cite{langgraph2024}, AutoGen \cite{wu2023autogen}, and CrewAI \cite{crewai2024} provide Python APIs for building multi-agent applications. They are excellent at composing agents within a single application, but they are not wire protocols: two systems built with different frameworks cannot interoperate without custom bridges, and none of them defines a cross-framework notion of intent, conflict, or resolution authority. MPAC sits one level below these frameworks---any of them could use MPAC as its coordination wire format.

\section{Design Goals and Non-Goals}
\label{sec:design}

MPAC's design is organized around six stated goals and an equally important set of explicit non-goals.

\subsection{Design Goals}

\begin{enumerate}[leftmargin=1.3em, itemsep=1pt, topsep=2pt]
    \item \textbf{Interoperability.} Different agent systems, written in different languages, by different organizations, should be able to coordinate through a shared message model.
    \item \textbf{Explicit coordination.} Agents should announce intent before acting whenever possible, so that overlaps can be detected proactively rather than discovered after the fact.
    \item \textbf{Causal traceability.} Every consequential action (commit, conflict report, resolution) should carry the causal context on which it was based, so that after-the-fact audits can reconstruct ``what each participant knew when.''
    \item \textbf{Structured conflict handling.} Conflicts should be first-class protocol objects with identity, category, severity, and attributable positions---not implicit failures or hidden race conditions.
    \item \textbf{Human-governed collaboration.} The protocol must support human override at every level. Automated resolution is a convenience; human authority is the fallback, and it must be reachable from any state.
    \item \textbf{Extensibility.} Optional features, implementation-specific extensions, and future message types must be possible without breaking core interoperability.
\end{enumerate}

\subsection{Non-Goals}

We emphasize the non-goals because they are as important as the goals for evaluating the design. MPAC is deliberately \emph{not}:

\begin{itemize}[leftmargin=1.3em, itemsep=1pt, topsep=2pt]
    \item A transport protocol. MPAC semantics must be realizable over WebSocket, HTTP, message queues, gRPC, or any other reliable message-passing substrate. The spec does not mandate a transport binding.
    \item A replacement for CRDTs, OT, or version control systems. It coordinates \emph{around} shared state, not the state itself.
    \item A single conflict-detection algorithm. It provides the structured representation of a conflict; the detection policy is left to coordinator implementations.
    \item A single security or trust framework. It defines three security profiles with different assumptions; concrete key management, identity issuance, and trust binding are out of scope.
    \item A replacement for MCP or A2A. MCP remains the right protocol for tool invocation; A2A remains the right protocol for single-principal delegation. MPAC complements both.
\end{itemize}

\subsection{Shared Principles}

Seven principles cut across all five protocol layers.

\paragraph{Intent before action.} In Governance-profile sessions, participants \emph{must} announce an \texttt{INTENT\_ANNOUNCE} before issuing \texttt{OP\_PROPOSE} or \texttt{OP\_COMMIT}. In Core-profile sessions, they \emph{should}. This single principle is what makes pre-emptive conflict detection possible.

\paragraph{Attributable actions.} Every operation, conflict report, and resolution must be attributable to a specific principal, so that audits can reconstruct responsibility across organizational boundaries.

\paragraph{Causal context.} \texttt{OP\_COMMIT}, \texttt{CONFLICT\_REPORT}, and \texttt{RESOLUTION} messages must carry a causal watermark in the envelope. Other messages should when available.

\paragraph{Human override.} The protocol must always expose an escalation path to a human principal for designated conflict classes. Human override is not an implementation afterthought; it is a normative requirement on governance-profile deployments.

\paragraph{Transport independence.} MPAC semantics must not depend on any specific transport. This is why the spec is organized around message types, state machines, and envelopes rather than connection lifecycles.

\paragraph{Algorithm independence.} A conflict object must look the same whether it was produced by a deterministic rule, a heuristic, a model inference, or a human review. This is what allows heterogeneous detection strategies to interoperate.

\paragraph{Coordinator-serialized total order.} MPAC does not provide linearizability in the strict distributed-systems sense; it provides a weaker but sufficient guarantee: the session coordinator serializes all state-mutating messages (commits, resolutions, intent-claim approvals) into a total order that participants eventually observe. This is analogous to single-leader replication.

\section{Protocol Model}
\label{sec:model}

\subsection{The Five Layers}

\begin{figure}[tb!]
\centering
\definecolor{sessionC}{HTML}{4C72B0}
\definecolor{intentC}{HTML}{55A868}
\definecolor{opC}{HTML}{DD8452}
\definecolor{conflictC}{HTML}{C44E52}
\definecolor{govC}{HTML}{8172B3}
\resizebox{\textwidth}{!}{%
\begin{tikzpicture}[
    font=\footnotesize,
    layer/.style={
        rectangle, rounded corners=2pt, draw=black!60, line width=0.4pt,
        minimum width=11cm, minimum height=0.95cm,
        inner sep=4pt, align=left
    },
    lbl/.style={font=\scriptsize\bfseries, text=white, anchor=west},
    desc/.style={font=\scriptsize, anchor=west, text=black!80},
    msgs/.style={font=\scriptsize\ttfamily, anchor=west, text=black!70},
    arr/.style={-{Latex[length=2mm]}, thick, black!55},
    oarr/.style={-{Latex[length=2mm]}, thick, govC, densely dashed},
    node distance=0.22cm
]

\node[layer, fill=sessionC!85] (s) at (0,0) {};
\node[layer, fill=intentC!85]  (i) [below=of s] {};
\node[layer, fill=opC!85]      (o) [below=of i] {};
\node[layer, fill=conflictC!85](c) [below=of o] {};
\node[layer, fill=govC!85]     (g) [below=of c] {};

\node[lbl] at ($(s.west)+(0.15,0.18)$) {1. SESSION};
\node[desc, text=white!92] at ($(s.west)+(2.15,0.18)$) {membership, identity, liveness};
\node[msgs, text=white!85] at ($(s.west)+(0.15,-0.18)$)
    {HELLO \,$\cdot$\, SESSION\_INFO \,$\cdot$\, HEARTBEAT \,$\cdot$\, COORDINATOR\_STATUS \,$\cdot$\, SESSION\_CLOSE};

\node[lbl] at ($(i.west)+(0.15,0.18)$) {2. INTENT};
\node[desc, text=white!92] at ($(i.west)+(2.15,0.18)$) {declare plans \emph{before} acting};
\node[msgs, text=white!85] at ($(i.west)+(0.15,-0.18)$)
    {INTENT\_ANNOUNCE \,$\cdot$\, INTENT\_UPDATE \,$\cdot$\, INTENT\_WITHDRAW \,$\cdot$\, INTENT\_CLAIM(\_STATUS)};

\node[lbl] at ($(o.west)+(0.15,0.18)$) {3. OPERATION};
\node[desc, text=white!92] at ($(o.west)+(2.35,0.18)$) {propose / commit mutations (OCC)};
\node[msgs, text=white!85] at ($(o.west)+(0.15,-0.18)$)
    {OP\_PROPOSE \,$\cdot$\, OP\_COMMIT \,$\cdot$\, OP\_REJECT \,$\cdot$\, OP\_SUPERSEDE \,$\cdot$\, OP\_BATCH\_COMMIT};

\node[lbl] at ($(c.west)+(0.15,0.18)$) {4. CONFLICT};
\node[desc, text=white!92] at ($(c.west)+(2.15,0.18)$) {structured, first-class disputes};
\node[msgs, text=white!85] at ($(c.west)+(0.15,-0.18)$)
    {CONFLICT\_REPORT \,$\cdot$\, CONFLICT\_ACK \,$\cdot$\, CONFLICT\_ESCALATE};

\node[lbl] at ($(g.west)+(0.15,0.18)$) {5. GOVERNANCE};
\node[desc, text=white!92] at ($(g.west)+(2.45,0.18)$) {authority, policy, human override};
\node[msgs, text=white!85] at ($(g.west)+(0.15,-0.18)$)
    {RESOLUTION (phase-scoped, authority-checked) \,$\cdot$\, role policy eval};

\draw[arr] ($(s.south west)+(0.45,0)$) -- ($(i.north west)+(0.45,0)$);
\draw[arr] ($(i.south west)+(0.45,0)$) -- ($(o.north west)+(0.45,0)$);
\draw[arr] ($(o.south west)+(0.45,0)$) -- ($(c.north west)+(0.45,0)$);
\draw[arr] ($(c.south west)+(0.45,0)$) -- ($(g.north west)+(0.45,0)$);

\draw[oarr] ($(g.east)+(-0.35,0.25)$) .. controls ($(g.east)+(0.9,0.25)$) and ($(i.east)+(0.9,0)$) .. ($(i.east)+(-0.35,-0.25)$);

\node[font=\tiny, govC, rotate=90, anchor=south] at ($(g.east)+(0.55,1.7)$) {override \& resume};

\node[font=\tiny, black!55, rotate=90, anchor=south] at ($(s.west)+(-0.15,-1.95)$) {normal lifecycle};

\node[draw=black!45, rounded corners=1pt, fill=black!4, minimum width=1.55cm, minimum height=5.1cm,
      font=\scriptsize, align=center, anchor=west]
      (coord) at ($(s.east)+(0.75,-2.05)$)
      {\textbf{Coordinator}\\[2pt]
       \scriptsize total order\\
       Lamport clock\\
       \texttt{epoch}\\
       snapshot \&\\ recovery\\
       scope overlap\\ detection\\
       authority\\ enforcement};

\foreach \n in {s,i,o,c,g}{
  \draw[black!35, line width=0.3pt] (\n.east) -- (\n.east-|coord.west);
}

\end{tikzpicture}%
}
\caption{MPAC's five logical coordination layers. The solid arrows on the left show the normal lifecycle: agents join a session, declare intents, propose or commit operations, and---if a conflict is detected---negotiate in the conflict layer, with the governance layer as the final authority. The dashed arrow on the right shows the governance override path: a \texttt{RESOLUTION} can unfreeze a contested scope and return control to the intent/operation layers. All layers share a single session coordinator that provides total order (Lamport clock + \texttt{coordinator\_epoch}), scope-overlap detection, snapshot-based fault recovery, and runtime enforcement of resolution authority. Implementations may merge these layers internally, but their externally visible semantics remain distinct.}
\label{fig:five-layers}
\end{figure}
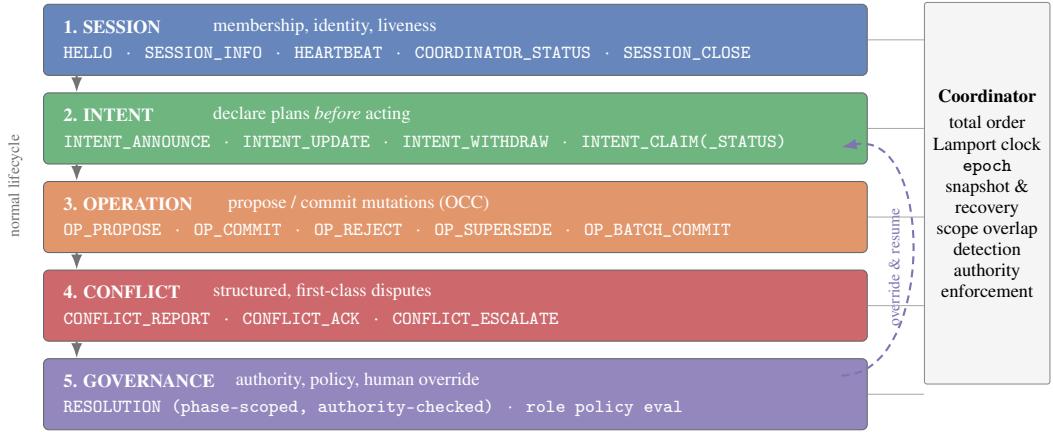

MPAC organizes coordination into five logical layers. Implementations may merge these layers internally, but the externally visible semantics must remain distinct.

\begin{enumerate}[leftmargin=1.3em, itemsep=1pt, topsep=2pt]
    \item \textbf{Session Layer.} Agents join a session, identify themselves, exchange credentials, negotiate capabilities and roles, and maintain liveness via heartbeats. Unregistered senders are rejected: every participant must complete a \texttt{HELLO} handshake before any other message type is accepted. This ``HELLO-first gate'' is normative and runtime-enforced in both reference implementations.
    \item \textbf{Intent Layer.} Agents declare what they \emph{plan} to do before doing it. An intent carries an objective, a scope (the set of resources it will touch), a priority, a time-to-live, and an optional basis (the causal context on which the plan was formed). Intents can be announced, updated, withdrawn, superseded, or claimed by a different agent if the owner becomes unavailable. Intent broadcasts give the coordinator a chance to detect overlaps \emph{before} any mutation occurs.
    \item \textbf{Operation Layer.} Agents propose and commit actual changes to shared state. Every commit carries \texttt{state\_ref\_before} and \texttt{state\_ref\_after} fields (SHA-256 hashes of the affected resources) for optimistic concurrency control. Stale commits are rejected with \texttt{STALE\_STATE\_REF}, and the proposing agent must rebase on the latest committed state and retry. Multi-resource changes use \texttt{OP\_BATCH\_COMMIT} with either \texttt{all\_or\_nothing} or \texttt{best\_effort} semantics.
    \item \textbf{Conflict Layer.} When the coordinator detects overlapping scopes or contradictory goals, it emits a \texttt{CONFLICT\_REPORT}---a structured object with identity, category (drawn from \texttt{scope\_overlap}, \texttt{concurrent\_write}, \texttt{semantic\_goal\_conflict}, \texttt{assumption\_contradiction}, \texttt{policy\_violation}, \texttt{authority\_conflict}, \texttt{dependency\_breakage}, \texttt{resource\_contention}; implementations may define additional categories), severity, and the set of implicated participants. Participants acknowledge with \texttt{CONFLICT\_ACK}, which can mark the position as ``seen,'' ``accepted,'' or ``disputed.'' Unresolved conflicts may be escalated via \texttt{CONFLICT\_ESCALATE}. If no resolution arrives within a timeout, the overlapping scope enters a \emph{frozen} state in which new mutations are blocked.
    \item \textbf{Governance Layer.} Authority rules---who can resolve what---are evaluated against the session's role policy (Section~23.1.5 of the specification). Only owners or designated arbiters may resolve conflicts pre-escalation; only the escalation target or an arbiter may resolve post-escalation. \texttt{RESOLUTION} messages carry the deciding principal's identity and the phase (pre- or post-escalation) at which the decision was made, so downstream audits can reconstruct the authority chain.
\end{enumerate}

\subsection{Two Execution Models}

MPAC defines two execution models that a session must declare up front in \texttt{SESSION\_INFO} and cannot mix.

\paragraph{Post-commit model.} The agent applies the mutation to shared state first, then declares the completed mutation via \texttt{OP\_COMMIT}. In this model, \texttt{OP\_COMMIT} is a \emph{notification} of a completed change. Conflicts discovered afterwards may require compensating operations. This model is suitable for Core-profile sessions and intra-team deployments where agents are trusted to act independently.

\paragraph{Pre-commit model.} The mutation is \emph{not} applied until the coordinator explicitly authorizes execution. The canonical flow is \texttt{OP\_PROPOSE} $\rightarrow$ coordinator review $\rightarrow$ authorization via \texttt{COORDINATOR\_STATUS} $\rightarrow$ proposer executes $\rightarrow$ \texttt{OP\_COMMIT}. Authorization alone does not transition the operation to \texttt{COMMITTED}; the proposer must later declare the executed mutation. This model requires Governance profile and is recommended for cross-organizational deployments where all changes must be reviewed before taking effect.

The explicit separation is deliberate: a protocol that silently ``depending on configuration'' flips between eager and lazy application is hard to reason about and hard to audit. A session must pick one model and stick with it.

\subsection{Consistency Model}

MPAC provides three distinct consistency regimes depending on coordinator availability.

\begin{enumerate}[leftmargin=1.3em, itemsep=1pt, topsep=2pt]
    \item \textbf{Coordinator-available (normal).} The coordinator serializes all state-mutating messages into a single total order. Participants eventually observe the same linearized sequence. The coordinator's Lamport clock is the authoritative ordering mechanism.
    \item \textbf{Coordinator-unavailable (degraded).} When participants detect coordinator unavailability, they must not perform state-mutating operations. Read-only and non-conflicting local work (planning, analysis) may continue, but no consistency guarantee applies to it.
    \item \textbf{Coordinator recovery (reconciliation).} After recovery, the coordinator rebuilds authoritative state from its latest snapshot plus audit log replay and bumps its \texttt{coordinator\_epoch}. If any participant violated rule (2) and mutated state during the outage, the divergence is detected when the participant's reported state fails to match the coordinator's recovered snapshot and the coordinator emits a \texttt{PROTOCOL\_ERROR} with \texttt{error\_code: STATE\_DIVERGENCE}; the situation must then be reconciled through governance-level resolution---the protocol deliberately does not auto-merge divergent states, because in a multi-principal setting silent auto-merge would violate attributability.
\end{enumerate}

This model is weaker than strict linearizability (participants observe changes with transport-dependent delay) but stronger than eventual consistency (there is always a single authoritative order at the coordinator). In distributed-systems terms, MPAC behaves like single-leader replication.

\subsection{Causal Watermarks}

Every consequential message carries a \texttt{Watermark} object in its envelope: a Lamport timestamp, a coordinator epoch, and optionally a sender-frontier summary. The Lamport clock guarantees that if event $A$ causally precedes event $B$, then $A$'s Lamport value is strictly less than $B$'s. The coordinator epoch distinguishes message sequences from different coordinator incarnations so that a post-recovery message cannot be mistakenly ordered against a pre-recovery one. Together these give auditors a total order over state-mutating events even when the transport delivers messages out of order.

Sender incarnation tracking adds a second guarantee: if an agent disconnects and reconnects, its new session is assigned a fresh incarnation ID, so replay of old messages from a crashed instance cannot be confused with new traffic.

\section{Messages and State Machines}
\label{sec:messages}

\subsection{The Twenty-One Message Types}

MPAC v0.1.13 defines 21 message types organized by layer. We list them here grouped by purpose; each has a dedicated JSON Schema and an \texttt{if/then} conditional constraint in the envelope schema.

\begin{table}[tb!]
\caption{The 21 MPAC message types grouped by protocol layer. All 21 have JSON Schema payload definitions and live Claude API demo coverage.}
\label{tab:messages}
\vspace{0.2em}
\begin{center}
\footnotesize
\begin{tabular}{lp{10.2cm}}
\toprule
\textbf{Layer} & \textbf{Message types} \\
\midrule
Session & \texttt{HELLO}, \texttt{SESSION\_INFO}, \texttt{HEARTBEAT}, \texttt{GOODBYE}, \texttt{SESSION\_CLOSE}, \texttt{COORDINATOR\_STATUS} \\
Intent  & \texttt{INTENT\_ANNOUNCE}, \texttt{INTENT\_UPDATE}, \texttt{INTENT\_WITHDRAW}, \texttt{INTENT\_CLAIM}, \texttt{INTENT\_CLAIM\_STATUS} \\
Operation & \texttt{OP\_PROPOSE}, \texttt{OP\_COMMIT}, \texttt{OP\_REJECT}, \texttt{OP\_SUPERSEDE}, \texttt{OP\_BATCH\_COMMIT} \\
Conflict  & \texttt{CONFLICT\_REPORT}, \texttt{CONFLICT\_ACK}, \texttt{CONFLICT\_ESCALATE}, \texttt{RESOLUTION} \\
Error & \texttt{PROTOCOL\_ERROR} \\
\bottomrule
\end{tabular}
\end{center}
\vspace{0.2em}
\end{table}

\paragraph{Session messages.} \texttt{HELLO} is always the first message from a participant; the coordinator responds with \texttt{SESSION\_INFO} declaring execution model, security profile, and the participant's granted roles. \texttt{HEARTBEAT} maintains liveness. \texttt{GOODBYE} lets a participant cleanly leave a session and declare how its remaining active intents should be handled (\texttt{withdraw}, \texttt{transfer}, or \texttt{expire}). \texttt{SESSION\_CLOSE} terminates a session and includes a summary record per Section 9.6.2 of the specification. \texttt{COORDINATOR\_STATUS} carries coordinator-originated notifications such as pre-commit authorizations, epoch changes, and health signals.

\paragraph{Intent messages.} \texttt{INTENT\_ANNOUNCE} declares a planned objective, scope, priority, and TTL. \texttt{INTENT\_UPDATE} modifies an active intent---for example, widening scope mid-plan, which may trigger new conflict detection. \texttt{INTENT\_WITHDRAW} cancels an intent voluntarily. \texttt{INTENT\_CLAIM} allows a surviving agent to take over another agent's suspended intent after a liveness timeout, subject to governance approval; \texttt{INTENT\_CLAIM\_STATUS} carries the approval or denial.

\paragraph{Operation messages.} \texttt{OP\_PROPOSE} requests authorization in pre-commit mode. \texttt{OP\_COMMIT} declares an executed mutation with \texttt{state\_ref\_before} and \texttt{state\_ref\_after}. \texttt{OP\_REJECT} is issued by the coordinator for validation failures, stale state refs, frozen-scope violations, or contested mutations. \texttt{OP\_SUPERSEDE} declares that a later operation replaces an earlier one in a supersession chain. \texttt{OP\_BATCH\_COMMIT} groups multiple operations with \texttt{all\_or\_nothing} or \texttt{best\_effort} semantics; failed \texttt{all\_or\_nothing} batches roll back all partially-registered operations.

\paragraph{Conflict messages.} \texttt{CONFLICT\_REPORT} is a coordinator-originated structured conflict object. \texttt{CONFLICT\_ACK} is how participants stake a position: its \texttt{ack\_type} is one of \texttt{seen}, \texttt{accepted}, or \texttt{disputed}. \texttt{CONFLICT\_ESCALATE} promotes an unresolved conflict to a designated arbiter. \texttt{RESOLUTION} declares a binding decision; the coordinator enforces that the sender has the authority to resolve at the conflict's current authority phase (pre- or post-escalation), so that only \texttt{owner}, \texttt{arbiter}, or coordinator-generated outcomes can bind a resolution.

\paragraph{Error message.} \texttt{PROTOCOL\_ERROR} is a lightweight message for signaling protocol-level problems that do not fit the conflict or operation-rejection categories: malformed payloads, invalid references, authorization failures, replay detection, stale state references, scope-frozen rejections, credential rejections, resolution timeouts, and state divergence between a participant and the coordinator's recovered snapshot. The coordinator's snapshot artifacts (including v0.1.13's replay-protection checkpoints so duplicate-message rejection survives restarts) are a persistence feature, not a wire message type, and do not appear in the message vocabulary.

\subsection{Three State Machines}

MPAC specifies three state machines with normative transition tables: Intent, Operation, and Conflict.

\paragraph{Intent state machine.} States: \texttt{ACTIVE}, \texttt{SUSPENDED} (owner liveness lost or departed with \texttt{intent\_disposition}: \texttt{transfer}), \texttt{TRANSFERRED}, \texttt{SUPERSEDED}, \texttt{WITHDRAWN}, \texttt{EXPIRED}. Transitions are driven by \texttt{INTENT\_ANNOUNCE}, \texttt{INTENT\_UPDATE}, \texttt{INTENT\_WITHDRAW}, \texttt{INTENT\_CLAIM\_STATUS}, liveness timeouts, and TTL expiry. A suspended intent transitions directly to \texttt{TRANSFERRED} upon an approved \texttt{INTENT\_CLAIM\_STATUS}; there is no intermediate claimed state. The spec enumerates every legal transition; the reference implementations enforce them.

\paragraph{Operation state machine.} States: \texttt{PROPOSED}, \texttt{COMMITTED}, \texttt{REJECTED}, \texttt{FROZEN} (referenced intent suspended), \texttt{ABANDONED} (sender unavailable), \texttt{SUPERSEDED}. In post-commit mode, operations enter the system directly as \texttt{COMMITTED} or \texttt{REJECTED}. In pre-commit mode, they traverse \texttt{PROPOSED} $\rightarrow$ \texttt{COMMITTED}, with coordinator authorization recorded as a flag \emph{within} \texttt{PROPOSED} rather than as a distinct state---authorization alone does not transition the operation to \texttt{COMMITTED}; only a subsequent \texttt{OP\_COMMIT} declaring the executed mutation does. \texttt{OP\_REJECT} is possible at any point up to commit.

\paragraph{Conflict state machine.} States: \texttt{OPEN}, \texttt{ACKED}, \texttt{ESCALATED}, \texttt{RESOLVED}, \texttt{CLOSED}, \texttt{DISMISSED}. \texttt{DISMISSED} covers both explicit dismissal and auto-dismissal when all related intents and operations have reached terminal states; \texttt{CLOSED} is the post-resolution archival terminal state. Scope freezing is a separate concept: when \texttt{resolution\_timeout\_sec} elapses without a \texttt{RESOLUTION}, the affected \emph{scope} enters a frozen state that blocks new mutations, but the underlying conflict itself remains in \texttt{OPEN} or \texttt{ACKED} until resolved, auto-dismissed, or force-closed by the three-phase frozen-scope degradation sequence. Frozen scopes are enforced at the \emph{target} level, not the intent level, so that omitting optional \texttt{intent\_id} fields cannot be used to bypass the freeze.

Cross-lifecycle rules, described as normative transition tables in Section 17 of the specification, govern how intent states interact with operation and conflict states---for example, withdrawing an intent automatically rejects any pending operations that referenced it (code \texttt{intent\_terminated}), and resolving a conflict unfreezes the associated scope.

\section{Security, Authorization, and Governance}
\label{sec:security}

\subsection{Three Security Profiles}

MPAC defines three security profiles with progressively stronger guarantees.

\begin{itemize}[leftmargin=1.3em, itemsep=1pt, topsep=2pt]
    \item \textbf{Open.} No credential required on \texttt{HELLO}. Suitable for intra-team sandboxes and testing.
    \item \textbf{Authenticated.} \texttt{HELLO} must carry one of five credential types (\texttt{bearer\_token}, \texttt{mtls\_fingerprint}, \texttt{api\_key}, \texttt{x509\_chain}, or \texttt{custom}). Replay protection is enforced: duplicate \texttt{message\_id} values are rejected with \texttt{REPLAY\_DETECTED}, and timestamps outside the replay window (recommended: 5 minutes) are rejected. Replay-protection state is persisted in the coordinator's periodic state snapshot so that rejection survives coordinator recovery.
    \item \textbf{Verified.} In addition to authenticated-profile guarantees, credentials must be verifiable against an issuer; self-asserted \texttt{arbiter} roles are rejected.
\end{itemize}

\subsection{Role Policy Evaluation}

When a participant sends \texttt{HELLO} with a set of requested roles, the coordinator evaluates them against the session's role policy (Section~23.1.5). Only authorized roles are granted; unauthorized roles are silently dropped, and a session with no policy declared cannot grant any role beyond the default \texttt{contributor}. This prevents a common category of bug in early drafts where an adversary could self-declare itself \texttt{arbiter} and use that role to rubber-stamp its own resolutions.

\subsection{Resolution Authority Enforcement}

The specification requires that only owners or arbiters may resolve conflicts before escalation, and only the designated escalation target or an arbiter may resolve after escalation. The reference implementations enforce this at the coordinator: \texttt{RESOLUTION} messages from principals without resolution authority are rejected with an explicit error code. This is not advisory---it is a runtime check, backed by adversarial tests that specifically try to bypass it.

\subsection{Frozen-Scope Enforcement}

Once a conflict's \texttt{resolution\_timeout\_sec} elapses without a resolution, the overlapping scope is frozen. Any subsequent \texttt{OP\_COMMIT} or \texttt{OP\_BATCH\_COMMIT} whose target set intersects the frozen scope is rejected with \texttt{SCOPE\_FROZEN}. New intents that are fully contained within a frozen scope are rejected at announcement time; intents that only partially overlap are accepted with a warning. This check is target-based---it does not rely on the intent's \texttt{intent\_id} field being present---so the freeze cannot be bypassed by omitting optional fields, a bypass we explicitly test against in the adversarial suite.

\subsection{Backend Health Monitoring}

Version 0.1.13 adds backend health monitoring primitives for production deployments. Coordinators expose liveness and readiness signals via \texttt{COORDINATOR\_STATUS}; participants can query them to decide whether to enter degraded mode. Snapshot recovery replays the audit log after loading the latest snapshot, bumps \texttt{coordinator\_epoch}, and republishes the new epoch to all reconnecting participants so that stale messages from the prior epoch are rejected.

\section{Reference Implementations}
\label{sec:refimpl}

\subsection{Two Independent Implementations, One Wire Format}

We release two reference implementations---one in Python and one in TypeScript---written independently against the specification. They are not two bindings of the same core; they are two separate implementations that must agree only on the wire. This is deliberate: having two independent implementations forces specification bugs, under-specified behavior, and hidden assumptions out into the open.

\begin{table}[tb!]
\caption{Reference implementation scale and test coverage as of v0.1.13.}
\label{tab:impl}
\vspace{0.2em}
\begin{center}
\footnotesize
\begin{tabular}{lccc}
\toprule
\textbf{Implementation} & \textbf{Source LOC} & \textbf{Test files} & \textbf{Test cases} \\
\midrule
Python  & $\sim$3{,}500 & 12 & 122 (including 34 adversarial) \\
TypeScript & $\sim$2{,}900 & 11 & 101 (including 32 adversarial) \\
\bottomrule
\end{tabular}
\end{center}
\vspace{0.2em}
\end{table}

\subsection{Cross-Language Interoperability}

A dedicated interoperability harness (\texttt{ref-impl/demo/run\_interop.sh}) exchanges 14 messages bidirectionally between the Python and TypeScript implementations: Python as coordinator with TypeScript as participant, and vice versa. The harness asserts byte-identical wire formats after normalization; it has zero deviation at v0.1.13. This is the strongest practical evidence that the specification is complete and unambiguous.

\subsection{Adversarial Testing}

Sixty-six adversarial tests (34 Python, 32 TypeScript) specifically target enforcement bypass attempts. These were written in response to actual findings from five rounds of independent audit on earlier versions. Categories include:

\begin{itemize}[leftmargin=1.3em, itemsep=1pt, topsep=2pt]
    \item \textbf{Unregistered sender attacks.} Sending any non-\texttt{HELLO} message before \texttt{HELLO} is rejected.
    \item \textbf{Credential bypass.} Authenticated/verified profiles reject \texttt{HELLO} without valid credentials.
    \item \textbf{Self-asserted arbiter.} Requesting the \texttt{arbiter} role when the session policy does not grant it is silently downgraded to the default role.
    \item \textbf{Unauthorized resolver.} \texttt{RESOLUTION} messages from principals without resolution authority at the current phase are rejected.
    \item \textbf{Frozen-scope evasion via omitted \texttt{intent\_id}.} Operations targeting a frozen scope are rejected even when the omitted optional field might otherwise bypass a naive check.
    \item \textbf{Snapshot recovery replay gap.} Replay protection state survives coordinator recovery via snapshot persistence.
    \item \textbf{Partial-overlap intent acceptance.} New intents that partially overlap a frozen scope are accepted with a warning; fully contained ones are rejected.
    \item \textbf{Batch atomicity rollback.} Failed \texttt{all\_or\_nothing} batches clean up all already-registered operations before returning the rejection.
\end{itemize}

\subsection{JSON Schema Conformance Closure}

All 21 message types have dedicated JSON Schema (Draft 2020-12) payload definitions in \texttt{ref-impl/schema/messages/}. The envelope schema in \texttt{envelope.schema.json} uses \texttt{if/then} conditional constraints to dispatch payload validation per \texttt{message\_type}. A third-party implementation can now achieve wire compatibility by JSON Schema validation alone---no prose reading required. Four shared object schemas (\texttt{Watermark}, \texttt{Scope}, \texttt{Basis}, \texttt{Outcome}) are used across multiple message payloads.

\section{Empirical Validation}
\label{sec:eval}

We validate MPAC through seven live multi-agent scenarios driven by the Anthropic Claude API \cite{anthropic2024claude}. Every scenario exercises real LLM decision-making; none uses mocked agents. All seven run over a WebSocket transport binding and together cover all 21 message types.

\subsection{Scenario Overview}

\begin{table}[tb!]
\caption{Seven distributed validation scenarios. Together they exercise all 21 message types under real Claude agent decision-making over a WebSocket transport binding.}
\label{tab:scenarios}
\vspace{0.2em}
\begin{center}
\footnotesize
\begin{tabular}{p{3.8cm}p{2.2cm}p{6.5cm}}
\toprule
\textbf{Scenario} & \textbf{Agents} & \textbf{Key protocol features exercised} \\
\midrule
Concurrent code editing & 2 & \texttt{INTENT\_ANNOUNCE}, scope-overlap \texttt{CONFLICT\_REPORT}, optimistic concurrency control via \texttt{state\_ref\_before}, \texttt{STALE\_STATE\_REF} rejection and rebase \\
Family trip planning & 3 & \texttt{task\_set} scope overlap on itinerary days and budget categories, natural-language \texttt{CONFLICT\_ACK} negotiation, atomic \texttt{OP\_BATCH\_COMMIT} \\
Overhead comparison & 3 & Back-to-back Traditional vs MPAC runs on the same workload with per-segment timing \\
Pre-commit + fault recovery & 3 & \texttt{OP\_PROPOSE} $\rightarrow$ authorization $\rightarrow$ \texttt{OP\_COMMIT}, \texttt{INTENT\_UPDATE}, \texttt{INTENT\_WITHDRAW} + \texttt{OP\_REJECT}, agent crash $\rightarrow$ liveness timeout $\rightarrow$ \texttt{INTENT\_CLAIM} with governance approval \\
Conflict escalation to arbiter & 3 & Disputed \texttt{CONFLICT\_ACK}, \texttt{CONFLICT\_ESCALATE}, Claude-powered arbiter \texttt{RESOLUTION}, multi-level governance authority chain \\
Interactive remote (pip-packaged) & 2+ & Two humans on different machines each give tasks to a local agent; coordinator runs on one side, WebSocket over LAN or ngrok \\
Cross-language interop & 2 & Python and TypeScript exchange 14 messages bidirectionally with zero wire deviation \\
\bottomrule
\end{tabular}
\end{center}
\vspace{0.2em}
\end{table}

Rather than reporting all seven in detail, we focus the remainder of this section on the three that carry the strongest empirical weight: the code-editing end-to-end run, the family-trip run (for domain generality), and the overhead comparison benchmark (for the performance claim).

\subsection{Code Editing with Optimistic Concurrency Control}

Two Claude agents---\emph{Alice} (security engineer persona) and \emph{Bob} (code quality engineer persona)---join a WebSocket coordinator that holds a Flask web application with five Python files containing intentional bugs: a token expiry bug in \texttt{auth.py}, N+1 queries in \texttt{models.py}, authentication duplication between \texttt{auth.py} and \texttt{auth\_middleware.py}, an unvalidated-input vulnerability in \texttt{api.py}, and an unclosed file handle in \texttt{utils.py}.

Each agent independently calls Claude to decide what to work on. Alice picks the token expiry bug in \texttt{auth.py} and \texttt{auth\_middleware.py}. Bob picks the duplication refactor, which touches \texttt{auth.py}, \texttt{auth\_middleware.py}, and \texttt{models.py}. Both announce their intents. The coordinator detects that the intents overlap on \texttt{auth.py} and \texttt{auth\_middleware.py} and emits a \texttt{CONFLICT\_REPORT} with category \texttt{scope\_overlap} and severity \texttt{medium}.

Both agents are asked, via independent Claude calls with no shared prompt or hardcoded priority, how to handle the conflict. Alice's position: ``This is a critical security vulnerability; I should proceed first and Bob can rebase his refactor onto my changes.'' Bob's position: ``Alice's security fix is urgent; my refactor can wait; let her go first.'' Two independent LLM calls reach the same conclusion through the protocol's structured conflict channel. The conflict is resolved. Alice commits her fix via \texttt{OP\_COMMIT}; Bob rebases on her committed \texttt{state\_ref\_after} and commits his refactor.

This scenario also exercises \texttt{STALE\_STATE\_REF} rejection: when Bob's initial commit attempt carries the original file hashes, the coordinator detects that Alice has since committed and returns \texttt{STALE\_STATE\_REF}. Bob fetches the new content, re-runs his refactor on top, and retries. The total message count for the full lifecycle---join, intent, conflict, resolution, two commits, session close---is modest (around 10--15 messages per agent in typical runs), and a full transcript is shipped with the repository as \texttt{ai\_demo\_transcript.json}.

The point of this scenario is not that ``agents can resolve conflicts politely.'' It is that the \emph{structured, auditable path} for them to do so is now a wire protocol, not ad-hoc prompting. A future agent with a different personality, a different LLM backend, or even a different organization can plug into the same coordinator and interoperate, because the semantics are in the protocol.

\subsection{Cross-Domain Generality: Family Trip Planning}

To argue that MPAC is a general coordination abstraction and not a code-editing framework in disguise, we run a second scenario that shares zero code with the first and lives in an entirely different domain: consumer trip planning.

Three Claude agents serve Dad, Mom, and Kid respectively, each with a distinct system prompt describing that family member's preferences, budget sensitivity, and authority level. They jointly plan a 5-day family trip. The shared state is an itinerary: five day-slots and several budget categories (lodging, food, activities, transport). Itinerary days and budget categories are modeled as \texttt{task\_set} scope resources.

Agents announce intents such as ``Dad: reserve Day 2 and Day 3 for camping, lodging category \textasciitilde \$0, activities category \$50 for gear rental.'' Mom's intent overlaps on Day 2 (``boutique minsu, lodging \$220''). The coordinator detects the scope overlap and emits a conflict report. Agents negotiate through structured \texttt{CONFLICT\_ACK} messages that carry natural-language justifications (``Kid's opinion: we camped last year, I'd like one fancy night''). A compromise emerges: Day 2 becomes boutique minsu, Day 3 stays camping. The itinerary is committed atomically via \texttt{OP\_BATCH\_COMMIT} with \texttt{all\_or\_nothing} semantics so that partial itineraries cannot be published.

The protocol primitives are unchanged from the code-editing scenario. Only the domain model (what counts as a ``resource,'' what counts as an ``objective'') differs. This is the minimum evidence for the claim that MPAC is a coordination \emph{layer}, not a software-development tool.

\subsection{Overhead Comparison Benchmark}

The third scenario is the one most directly relevant to the performance claim in the abstract. We designed it to answer a specific question: \emph{does MPAC actually make multi-agent coordination faster, or does it just move the same latency around?}

\paragraph{Setup.} Three Claude agents---Alice, Bob, and Charlie---review a cross-module pull request that touches three modules with known interactions. Each agent is responsible for reviewing its own module but must coordinate with the others because the modules have cross-cutting concerns. We run the exact same scenario twice, back-to-back, using the same Claude model (\texttt{claude-sonnet-4}) and the same prompts:

\begin{enumerate}[leftmargin=1.3em, itemsep=1pt, topsep=2pt]
    \item \textbf{Traditional mode (baseline).} Agents review serially, human-mediated: each one waits for the previous one's output, clarification questions cause round-trip delays, and conflicts are discovered and resolved \emph{after} the work is done (post-hoc rework).
    \item \textbf{MPAC mode.} Agents review in parallel over a WebSocket coordinator. \texttt{INTENT\_ANNOUNCE} surfaces scope overlaps \emph{before} any review work begins; structured \texttt{CONFLICT\_ACK} lets agents exchange positions without human mediation; commits proceed in parallel.
\end{enumerate}

Every segment of every run is timed with per-segment wall-clock instrumentation and classified as either \emph{decision time} (the agent is actively thinking, i.e., waiting on a Claude API response) or \emph{coordination overhead} (the agent is blocked waiting for another agent, a round-trip, or post-hoc rework).

\paragraph{Results.} Table~\ref{tab:overhead} shows the results of a single controlled run.

\begin{table}[tb!]
\caption{
Overhead comparison on a 3-agent cross-module PR review. 
Wall-clock time is the externally observed elapsed real time from session start to the last commit; decision time is the cumulative time agents spend waiting on Claude API responses; coordination overhead is the cumulative time agents spend blocked on serialization, round-trips, or post-hoc rework. In Traditional mode wall-clock equals decision + overhead because the three agents are serialized (63.11 + 68.65 = 131.76 s); in MPAC mode wall-clock is much smaller than decision + overhead because the agents run in parallel and wall-clock is bounded by the slowest single agent rather than by the sum across agents (27.38 s vs. 57.13 + 3.02 = 60.15 s of cumulative work). 
One controlled run; see Section~\ref{sec:limitations} for discussion of statistical scope.}
\label{tab:overhead}
\vspace{0.2em}
\begin{center}
\footnotesize
\begin{tabular}{lccc}
\toprule
\textbf{Metric} & \textbf{Traditional} & \textbf{MPAC} & \textbf{Change} \\
\midrule
Wall-clock time (s)               & 131.76 & 27.38 & $-79.2\%$ (4.8$\times$ speedup) \\
\hline
Decision time (s)                 & 63.11  & 57.13 & $-9.5\%$ (preserved) \\
Coordination overhead (s)         & 68.65  & 3.02  & $-95.6\%$ \\
Overhead as \% of wall-clock      & 52.1\% & 11.0\% & $-41.1$ pp \\
\bottomrule
\end{tabular}
\end{center}
\vspace{0.2em}
\end{table}

\paragraph{Interpretation.} Two observations matter here, and they matter independently.

\emph{First,} coordination overhead drops from 68.65\,s to 3.02\,s---a 95.6\% reduction, or absolute savings of 65.6\,s. This is the effect of replacing serialized human-mediated coordination with protocol-level intent broadcast, pre-emptive conflict detection, and structured parallel negotiation. It is the core performance claim of the paper, and it is exactly the kind of saving a coordination protocol should produce.

\emph{Second,} and equally important, per-agent decision time is roughly preserved (63.11\,s $\rightarrow$ 57.13\,s, a 9.5\% change, most of which is within run-to-run LLM variance). \textbf{This is the load-bearing methodological point.} A trivial ``optimization'' could have come from cutting prompts shorter or skipping review steps, which would have shown up as reduced decision time. The fact that decision time is essentially unchanged is the strongest evidence that MPAC is not compressing the \emph{work}; it is eliminating the \emph{waiting}. We argue that this is exactly the right factorization for a coordination protocol: the decision time is whatever the underlying model and task demand, and the protocol's job is to drive the coordination overhead toward zero.

\emph{Third,} the end-to-end wall-clock speedup is 4.8$\times$. For a cluster of agents running at $O(10)$ seconds per decision, reducing the coordination overhead is the difference between a flow that finishes in under half a minute and one that drags past two minutes.

We are deliberately conservative about generalizing from a single run; Section~\ref{sec:limitations} discusses the statistical scope of this result.

\subsection{Additional Scenarios}

Three further scenarios round out coverage of the protocol surface. The \emph{pre-commit + fault recovery} scenario exercises the pre-commit execution model: \texttt{OP\_PROPOSE} $\rightarrow$ coordinator authorization $\rightarrow$ \texttt{OP\_COMMIT}. It also simulates an agent crash: one agent becomes unresponsive, liveness tracking flags it, its in-flight intents are suspended, and a surviving agent successfully claims the suspended work via \texttt{INTENT\_CLAIM} with governance approval. The \emph{escalation} scenario exercises multi-level governance: two owner agents dispute a scope overlap, both mark their positions as ``disputed,'' one escalates to a designated arbiter, and the arbiter (a third Claude agent with an explicit judicial system prompt) issues a binding \texttt{RESOLUTION}. Finally, the \emph{remote pip-packaged} scenario validates cross-host deployment: two users on different machines, each running a local Claude agent, join a shared coordinator over WebSocket (LAN or ngrok) and see real-time notifications when the other agent commits.

Together with the cross-language interop test, these seven scenarios cover all 21 message types at least once under live LLM decision-making---giving the protocol, the schemas, and the reference implementations end-to-end empirical support.

\section{Related Work}
\label{sec:related}

\paragraph{Agent protocols.} MCP \cite{anthropic2024mcp} standardizes agent-to-tool interaction; A2A \cite{google2024a2a} standardizes single-principal agent-to-agent delegation. MPAC is complementary: it addresses multi-principal coordination, which neither protocol targets. A comparison framework for agent communication protocols appears in \cite{ehtesham2025survey}, which explicitly identifies multi-principal coordination as an open gap.

\paragraph{Multi-agent frameworks.} LangGraph \cite{langgraph2024}, AutoGen \cite{wu2023autogen}, CrewAI \cite{crewai2024}, and similar frameworks provide in-process composition of agents. They are libraries, not wire protocols, and do not define cross-framework interoperability. MPAC could serve as the wire format beneath any of them.

\paragraph{Distributed systems antecedents.} MPAC's causal watermark design draws directly from Lamport's logical clocks \cite{lamport1978time} and the single-leader replication pattern familiar from consensus protocols such as Paxos \cite{lamport1998paxos} and Raft \cite{ongaro2014raft}. Unlike these, MPAC does not solve consensus under Byzantine or arbitrary failures---it assumes a well-behaved coordinator and is instead focused on semantic-layer coordination (intent, conflict, governance) above a reliable message-passing substrate. CRDTs \cite{shapiro2011crdt} and OT \cite{ellis1989concurrency} handle concurrent edits to shared data structures; MPAC addresses coordination \emph{around} shared state and can sit alongside a CRDT-backed store without modification.

\paragraph{Classical multi-agent systems and argumentation.} The broader multi-agent systems literature \cite{wooldridge2009multiagent} includes decades of work on agent communication languages (KQML, FIPA-ACL \cite{fipa2002acl}), argumentation frameworks, contract nets, and negotiation protocols. MPAC borrows conceptual elements---especially the insistence that ``disagreement'' should be a first-class object---but targets the narrower and more concrete problem of coordinating present-day LLM agents over shared mutable state, with machine-enforceable wire semantics and two independent reference implementations.

\paragraph{Version control and collaborative editing.} Git, Mercurial, and collaborative editing systems such as Google Docs solve related problems for human users: they make concurrent edits to shared state manageable through merge semantics or OT. MPAC differs in that its participants are autonomous agents whose decisions must be \emph{pre-announced} (so that overlap detection is pre-emptive rather than post-hoc) and \emph{attributed} (so that multi-principal audit is possible). A human using Git does not need to announce their intent to edit a file; an autonomous agent acting on behalf of one principal among many should.

\section{Limitations and Threats to Validity}
\label{sec:limitations}

We are deliberately explicit about the limitations of this work, because honesty about what has and has not been demonstrated is the right posture for a draft protocol.

\paragraph{Single-run benchmark.} The overhead comparison in Table~\ref{tab:overhead} reports a single controlled run of a single 3-agent workload. The magnitudes we report---$95.6\%$ overhead reduction, $4.8\times$ wall-clock speedup---are not asymptotic claims. They are an \emph{existence demonstration} that the overhead delta is large and specific enough to be worth reporting, against a controlled baseline on the same prompts and model. A proper benchmark study would vary agent count, task difficulty, per-decision Claude latency, and conflict density; we plan to publish such a study as follow-up work. We encourage readers to treat the present numbers as calibration, not as a performance curve.

\paragraph{Coordinator is a single point of failure.} The reference implementations use a single coordinator process. Coordinator epoch fencing (Section~8.1.1.4 of the specification) is implemented and gives the protocol a basis for future multi-coordinator handover, and snapshot-based fault recovery is implemented and tested, but split-brain detection across concurrent coordinator instances is not yet exercised. A full multi-coordinator handover story---with cross-instance fencing and liveness---remains future work.

\paragraph{No formal verification.} The three state machines have normative transition tables in the specification and are exercised by adversarial tests in both implementations, but they have not been formally verified (e.g., in TLA+). We are particularly interested in cross-lifecycle interactions between Intent, Operation, and Conflict state machines, where edge cases are most likely to hide.

\paragraph{No head-to-head comparison with frameworks.} We do not directly compare MPAC against LangGraph-, AutoGen-, or CrewAI-based multi-agent implementations of the same workloads. This is partly philosophical---MPAC is a protocol, not a framework, and a fair comparison would require building a protocol-free baseline of comparable engineering maturity---and partly practical, because such a comparison is a substantial study in its own right. We flag this as important follow-up work.

\paragraph{Byzantine assumptions.} The current specification assumes that the coordinator is well-behaved (crash-stop, not Byzantine) and that participants follow the protocol unless they are explicitly testing enforcement. A fully multi-principal adversarial setting---where one participant is actively trying to lie about state, forge causality, or evade authority checks---is partially addressed by the verified security profile and adversarial tests, but a rigorous threat model is future work. Signature verification and trust binding across organizational boundaries remain gaps.

\paragraph{Scope detection is coordinator-local.} The current coordinator detects scope overlaps by intersecting declared resource sets. Richer semantics---e.g., ``Alice's refactor will eventually touch all callers of this function''---require richer scope languages than \texttt{file\_set} and \texttt{task\_set}. The specification's \texttt{Scope} object is extensible for this purpose, but we have not yet validated richer scope kinds in the reference implementations.

\paragraph{LLM variance.} All seven validation scenarios use the Claude API for agent decisions. LLMs are stochastic, and individual runs vary. Our protocol-level claims---message types, state machines, conflict structure---do not depend on LLM determinism, but the narrative quality of the negotiation transcripts (``agents reached the same conclusion independently'') does. We are not claiming that every run produces the same negotiation path; we are claiming that the protocol provides the structured channel through which any negotiation path, deterministic or not, becomes auditable.

\section{Discussion and Future Work}
\label{sec:discussion}

\paragraph{Where MPAC sits in the stack.} If MCP is the ``transport layer'' of agent-tool interaction and A2A is a ``single-principal orchestration layer,'' MPAC is a ``multi-principal coordination layer'' that sits above both and beside neither. An agent in a realistic future deployment will likely use MCP to call tools, A2A to delegate to its own sub-agents, and MPAC to coordinate with agents belonging to other principals---all three at once, at different layers of the same stack.

\paragraph{From v0.1 to v0.2.} The roadmap for v0.2.0 includes richer scope expressiveness, post-commit rollback semantics, cross-session coordination, a compact binary envelope, and scope-based subscription (so that participants need not receive every message). These additions are evolution, not redesign---the five-layer model and 21-message vocabulary have stabilized across thirteen revision rounds, and further audit feedback will inform incremental changes rather than structural ones.

\paragraph{Conformance harness.} A next priority is an automated conformance test suite that any third-party implementation can run against its own wire output. The JSON Schema conformance closure in v0.1.13 is the foundation; the harness would add a standard set of interop traces (e.g., the 14-message interop test generalized to cover every message type) plus negative test cases for the adversarial categories in Section~\ref{sec:refimpl}.

\paragraph{Formal verification.} TLA+ modeling of the three state machines, especially the cross-lifecycle interactions, is the next step we expect to produce the highest marginal confidence. Early versions of the specification contained silent cross-lifecycle bugs that took multiple audit rounds to surface; formal modeling would catch the next such class of bugs before they reach the reference implementations.

\paragraph{Adoption path.} We are not asking readers to adopt MPAC in production today. We are asking them to read the specification, try the reference implementations, run the seven demos, and tell us where the design is wrong. The protocol is in a draft state specifically so that feedback from external implementers can still reshape it. An open specification with two reference implementations and seven live demos is our best attempt to lower the cost of that feedback.

\section{Conclusion}
\label{sec:conclusion}

MPAC is a protocol-layer answer to a concrete question: \emph{when agents serving different principals need to work together, how should they coordinate?} The question is not answered by MCP (wrong layer) or A2A (wrong principal model), and it is not answered well by layering message queues, locks, or in-process frameworks on top of those protocols. MPAC's answer is a five-layer model---Session, Intent, Operation, Conflict, Governance---with 21 message types, three state machines, Lamport-clock causality, two execution models, three security profiles, and machine-enforceable conformance via JSON Schema.

We release two fully interoperable reference implementations, 223 tests including 66 adversarial tests, seven live Claude-driven multi-agent demos covering all 21 message types, and a controlled benchmark showing that structured pre-announcement of intent eliminates 95.6\% of coordination overhead and produces a 4.8$\times$ wall-clock speedup on a 3-agent cross-module code-review task---while preserving per-agent decision time.

The specification, implementations, schemas, and demo transcripts are open source. We invite the multi-agent systems community---both researchers and practitioners building on MCP, A2A, and adjacent protocols---to read, break, and improve the design.

\paragraph{Code and artifacts.} The full protocol specification (\texttt{SPEC.md}), developer reference, JSON Schema suite, Python and TypeScript reference implementations, distributed demo transcripts, and version history are released in the project repository. The pip-installable \texttt{mpac\_protocol} package and the \texttt{mpac-starter-kit} archive let two users on different machines run collaborative multi-agent sessions with a single command per side.

\bibliographystyle{plain}
\bibliography{sample-base}

\end{document}